\documentstyle[sprocl]{article}

\def\Journal#1#2#3#4{{#1} {\bf #2}, #3 (#4)}

\def\NPB{{\em Nucl. Phys.} B}
\def\NPA{{\em Nucl. Phys.} A}
\def\PLB{{\em Phys. Lett.}  B}

\def\PRD{{\em Phys. Rev.} D}
\def\PRC{{\em Phys. Rev.} C}

\def\be{\begin{equation}}
\def\ee{\end{equation}}
\def\bea{\begin{eqnarray}}
\def\eea{\end{eqnarray}}
\begin{document}

\title{ About QCD sum rules for pion-nucleon coupling}

\author{Hungchong Kim}

\address{ Department of Physics, Tokyo Institute of Technology, Tokyo
152-8551,
Japan
\\E-mail: hckim@th.phys.titech.ac.jp}

\maketitle\abstract{
We point out subtleties in the previous sum rule calculations of
the $\pi NN$ coupling  
associated with using either the  
PV or the PS coupling schemes in modeling the  phenomenological side.
We propose a sum rule which is independent of the coupling schemes used and 
has less uncertainty in the OPE.  The obtained value is 
$g_{\pi N}=9.76 \pm 2.04$ where the uncertainty mainly comes
from the quark-gluon mixed condensate.}

%\section{Guidelines}
\vspace{10pt}

Since first introduced by Shifman, Vainshtein and Zakharov~\cite{SVZ},
QCD sum rule has been widely used to study the hadron properties such as
masses or couplings of baryons~\cite{qsr}. 
QCD sum rule is a framework which connects the physical parameters with
QCD parameters.  In this framework, a correlation function is 
introduced in terms of  interpolating fields constructed from quark and
gluon fields.
Then, the correlation function, 
on the one hand,  is calculated by Wilson's
operator product expansion (OPE) and, on the other hand, its  phenomenological
``ansatz'' is constructed.  A physical quantity of
interest is extracted by matching the two descriptions in the
deep Euclidean region ($q^2 = - \infty$) making use of the dispersion relation.
The extracted value therefore should be independent of possible ansatz
in order to be physically meaningful.

One important quantity to be determined in hadron physics is the pion-nucleon
coupling constant, $g_{\pi N}$, whose empirical value is known to be around 
13.4 but it is of interest
to determine the coupling from QCD.
QCD sum rule can be used for such purpose and indeed there are such 
calculations of $g_{\pi N}$~\cite{qsr,hat,krippa}.  
Reinders, Rubinstein and Yazaki~\cite{qsr} calculated
$g_{\pi N}$ by retaining only the first nonperturbative term in 
the OPE.  Later Shiomi and Hatsuda (SH)~\cite{hat} improved
the calculation by including higher order terms in the OPE. SH considered
the two-point correlation function for the nucleon interpolating field
$J_N$,
\begin{eqnarray}
\Pi (q, p) = i \int d^4 x e^{i q \cdot x} \langle 0 | T[J_N (x) 
{\bar J}_N (0)]| \pi (p) \rangle \ ,
\label{two}
\end{eqnarray}
and evaluated the OPE in the soft-pion limit ($p_\mu \rightarrow 0$).
More recently, Birse and Krippa (BK)~\cite{krippa}  pointed out that
the use of the soft-pion limit does not constitute an independent sum
rule from the nucleon sum rule and they considered
the sum rule beyond the soft-pion limit.

In both sum rules, the phenomenological side
of the correlator is  constructed by 
using the PS interaction Lagrangian,
\begin{eqnarray}
{\cal L}_{ps} = g_{\pi N} {\bar \psi} i \gamma_5 {\vec \tau}
\cdot {\vec \pi} \psi\ .
\label{ps}
\end{eqnarray}
However,  the PV interaction Lagrangian,
\begin{eqnarray}
{\cal L}_{pv} = {g_{\pi N} \over 2m}  {\bar \psi} \gamma_5 \gamma_\mu
 {\vec \tau} \cdot \partial^\mu {\vec \pi} 
\psi\ ,
\label{pv}
\end{eqnarray}
can also be used which could 
change the results. This is because, in QCD sum rules,
on-shell quantities are extracted from far Euclidean region where
descriptions from the two Lagrangians are different.  

To achieve the independence from the PS and PV coupling schemes, 
it is natural to go beyond the soft-pion limit.  Then the method
proposed by Ref.~\cite{krippa} seems to be appropriate.
The $\gamma_5 \not\!p$ structure considered in Ref.~\cite{krippa}
uses  the PS Lagrangion,
\begin{eqnarray}
-{i g_{\pi N} \lambda_N^2  m  \over 
(q^2 - m^2 +i \epsilon)[(q-p)^2 - m^2 + i \epsilon]} + \cdot \cdot \cdot\ . 
\end{eqnarray}
Then after some manipulation, the sum rule formula can
be succinctly casted as 
\begin{eqnarray}
g_{\pi N} + A M^2 = f(M)\ .
\label{ksum}
\end{eqnarray}
$f(M)$ is a  function of the Borel mass $M$. The unknown simple pole 
contribution $A$ is associated with the transition 
$N \rightarrow N^*$~\cite{ioffe1}  and  estimated to be small~\cite{krippa}.  
With the PV scheme,
the phenomenological side takes the form,
\begin{eqnarray}
-{i g_{\pi N} \lambda_N^2 / 2 m \over (q-p)^2 - m^2 + i \epsilon }
-{i g_{\pi N} \lambda_N^2  m \over 
(q^2 - m^2 +i \epsilon)[(q-p)^2 - m^2 + i \epsilon] } + \cdot \cdot \cdot\ .
\end{eqnarray}
Note that the double pole is the same as above but 
there is an additional simple pole of $N\rightarrow N$.  
It means that physical content of
the single pole  is coupling scheme-dependent.  
If this were used in BK's sum rule, then instead of Eq.~(\ref{ksum}),
BK  would have obtained  
\begin{eqnarray}
g_{\pi N} \left (1 - {M^2 \over 2 m^2} \right ) + A M^2 = f(M)\ .
\label{ksum2}
\end{eqnarray}
Around  $M \sim m \sim 1$ GeV, the  new term is  0.5, impling
that $A$ is large within the prescriptions used in Ref.~\cite{krippa}.

One more Dirac structure, $\gamma_5 \sigma_{\mu\nu}$, contains the
double pole only, independent of the coupling scheme.
The common phenomenological side for this structure is given by
\begin{eqnarray}
-{ g_{\pi N} \lambda_N^2  p^\mu q^\nu  \over 
(q^2 - m^2 +i \epsilon)[(q-p)^2 - m^2 + i \epsilon]} + \cdot \cdot \cdot\ . 
\end{eqnarray}
This structure is zero in the soft-pion limit, but beyond the
soft-pion limit, this can provide a new sum rule 
for $g_{\pi N}$. Since there is no simple 
nucleon pole in this case, the simple pole structure comes only from 
$N \rightarrow  N^*$.

To construct a sum rule for  $\gamma_5 \sigma^{\mu\nu} p_\mu q_\nu$ structure,
we consider the correlation function
\begin{eqnarray}
\Pi (q,p) = i \int d^4 x e^{i q \cdot x} \langle 0 | T[J_p (x) 
{\bar J}_n (0)]| \pi^+ (p) \rangle\ .
\label{two2}
\end{eqnarray}
Here $J_p$ is the proton interpolating field of Ioffe~\cite{ioffe2},
\begin{eqnarray}
J_p = \epsilon_{abc} [ u_a^T C \gamma_\mu u_b ] \gamma_5 \gamma^\mu d_c
\end{eqnarray}
and the neutron interpolating field $J_n$ is obtained by replacing
$(u,d) \rightarrow (d,u)$.  
In the OPE,  we will only keep the diquark component of the 
pion wave function and use the vacuum saturation hypothesis
to factor out 
higher dimensional operators in terms of the pion wave function and the 
vacuum expectation value.

The calculation of the correlator, Eq.~(\ref{two2}), in the coordinate
space contains the following quark-antiquark component of the pion wave 
function, 
\begin{eqnarray}
&&D^{\alpha\beta}_{a a'} \equiv 
\langle 0 | u^\alpha_a (x) {\bar d}^\beta_{a'} (0) | \pi^+ (p) \rangle
\nonumber\\ 
%\end{eqnarray}
%This matrix element can be written in terms of three
%Dirac structures,
%\begin{eqnarray}
&&= {\delta_{a a'}\over 12} \Big [ 
(\gamma^\mu \gamma_5)^{\alpha \beta}
\langle 0 |  
{\bar d} (0) \gamma_\mu \gamma_5  u (x) | \pi^+ (p) \rangle\ 
+  (i \gamma_5)^{\alpha \beta} 
\langle 0 |  
{\bar d}(0) i \gamma_5  u (x) | \pi^+ (p) \rangle\nonumber \\ 
&& - {(\gamma_5 \sigma^{\mu\nu})^{\alpha\beta}\over 2}
\langle 0 |  
{\bar d}(0) \gamma_5 \sigma_{\mu\nu}  u (x) | \pi^+ (p) \rangle \Big ]\ .
\label{dd}
\end{eqnarray}
These matrix elements can be written in terms of pion wave 
function.  
Since we are doing the calculation up to the first order of $p_\mu q_\nu$,
we need only the overall normalization of the wave functions. 
In fact, to leading order in the pion momentum, 
the first and third matrix elements are given as~\cite{bely},
\begin{eqnarray}
\langle 0 | {\bar d} (0) \gamma_\mu \gamma_5  u (x) | \pi^+ (p) \rangle\ 
&=& i \sqrt{2} f_\pi p_\mu + O(x^2) \label{d1} \\
\langle 0 |{\bar d}(0) \gamma_5 \sigma_{\mu\nu}  u (x) | \pi^+ (p) \rangle\ 
&=&i \sqrt{2} (p_\mu x_\nu - p_\nu x_\mu) 
{f_\pi m_\pi^2 \over 6 (m_u + m_d)}\ .
\label{d2}
\end{eqnarray}
In Eq.~(\ref{d1}), $O(x^2)$ contains the twist 4 term 
which 
does not contribute to our sum rule up to the dimension we consider below.
Note that in Eq.~(\ref{d2}) the factor $f_\pi m_\pi^2 / (m_u + m_d)$ can
be written as $-\langle {\bar q} q \rangle / f_\pi$ by making use of
the Gell-Mann$-$Oakes$-$Renner relation.

In our sum rule for the $\gamma_5 \sigma^{\mu\nu} p_\mu q_\nu$ structure,
the second matrix element in Eq.~(\ref{dd}) does not contribute 
up to dimension 7. 
It is straightforward to calculate the OPE. 
We match the OPE with its phenomenological counterpart and take the 
Borel transformation. Then the unknown parameter, $\lambda_N$, which
represents the coupling strength of the interpolating field to the physical
nucleon, is eliminated by using the nucleon chiral odd sum rule. 
For more discussion, see Ref.~\cite{hung}.
The final expression for the sum rule is
\begin{eqnarray}
&&g_{\pi N} (B M^2 + 1)    \left ({f_\pi \over m}\right ) \\
&& =
{M^4 E_0 (x_\pi)/3 + 16 \pi^2 f_\pi^2 M^2/3 + 
\pi^2 \left \langle {\alpha_s \over \pi} {\cal G}^2 \right \rangle/54 
-2\pi^2 m_0^2 f_\pi^2 /3 
\over M^4 E_1(x_N) - 
\pi^2 \left \langle {\alpha_s \over \pi} {\cal G}^2 \right \rangle/6 }
\nonumber\ .
\label{oursum}
\end{eqnarray}
Here, $B$ represents the unknown simple pole of $N\rightarrow N^*$,
$x_\pi = S_{\pi N}/M^2$ with $S_{\pi N}$ being the continuum threshold,
$x_N$ is corresponding one for  the nucleon sum rule,
and $E_n(x) = 1 - (1 + x + \cdot\cdot\cdot + x^n/n!) e^{-x}$.
In this sum rule, the main  source for the error comes from  $m_0^2$
which is related
to quark-gluon mixed condensate via
$ \langle {\bar q} g_s \sigma \cdot {\cal G} q \rangle
=m_0^2 \langle {\bar q} q \rangle $\ .  
This contributes only to the highest dimensional term in the OPE 
so this will be  suppressed in the
Borel window chosen.
According to Ref.~\cite{Ovc},
we take the range, $0.6 \le m_0^2 \le 1.4$ GeV$^2$ and 
see the sensitivity of our results. 
 
We fit the RHS of Eq.~(\ref{oursum}) with a straight line within the
Borel window chosen appropriately.
Our result is  $g_{\pi N} = 9.76 \pm 2.04$,
where the error comes from the uncertainty in $m_0^2$.
This result is independent of the coupling schemes and its error is
relatively small.

%\vspace{5pt}
\section*{Acknowledgments}
This work is supported by Research Fellowship of the Japan
Society for the promotion of Science.


\section*{References}
\begin{thebibliography}{99}


\bibitem{SVZ}     M.A. Shifman, A.I. Vainshtein, and V.I. Zakharov,
                            \Journal{\NPB} {147} {385,448} {1979}.
\bibitem{qsr}      L.J. Reinders, H. Rubinstein and
                    S. Yazaki, {\it Phys. Rep.} {127}{1} {1985}.
\bibitem{hat}     H. Shiomi and T. Hatsuda,
                            \Journal{\NPA} {594}{294} {1995}.
\bibitem{krippa}     M. C. Birse and B. Krippa,
                            \Journal{\PLB}{373}{9} {1996};
    M. C. Birse and B. Krippa,
                            \Journal{\PRC}{54}{3240} {1996}.
\bibitem{ioffe1}     B. L. Ioffe and A. V. Smilga,
                            \Journal{\NPB} {234}{109} {1984}.
\bibitem{ioffe2}     B. L. Ioffe,
                            \Journal{\NPB} {188}{317} {1981}.
 
\bibitem{bely}     V. M. Belyaev, V. M. Braun, A. Khodjamirian and R. R\"uckl,
                            \Journal{\PRD}{51}{6177} {1995}.
\bibitem{Ovc}       A. A. Ovchinnikov and A. A. Pivovarov,
                            {\it Yad. Fiz.} {48}{1135} {1988};
    M. Kremer and G. Schierholz,
                            \Journal{\PLB}{194}{283} {1987};
    M. V. Polyakov and C. Weiss,
                            \Journal{\PLB}{387}{841} {1996}.
\bibitem{hung}    Hungchong Kim, Su Houng Lee and Makoto Oka,
                            {\it Los Alamos Preprint nucl-th/9809004,
                         Submitted to Physics Letters B}.


\end{thebibliography}
\end{document}